\documentclass[twocolumn,prd,aps,epsfig,floats,showpacs]{revtex4}
\def\DESepsf(#1 width #2){\epsfxsize=#2 \epsfbox{#1}}
\usepackage{epsfig}
\usepackage{graphicx}
\def\bmatrix{\left[\begin{array}}
\def\ematrix{\end{array}\right]}

\begin{document}

%

\let\a=\alpha      \let\b=\beta       \let\c=\chi        \let\d=\delta
\let\e=\varepsilon \let\f=\varphi     \let\g=\gamma      \let\h=\eta
\let\k=\kappa      \let\l=\lambda     \let\m=\mu
\let\o=\omega      \let\r=\varrho     \let\s=\sigma
\let\t=\tau        \let\th=\vartheta  \let\y=\upsilon    \let\x=\xi
\let\z=\zeta       \let\io=\iota      \let\vp=\varpi     \let\ro=\rho
\let\ph=\phi       \let\ep=\epsilon   \let\te=\theta
\let\n=\nu
\let\D=\Delta   \let\F=\Phi    \let\G=\Gamma  \let\L=\Lambda
\let\O=\Omega   \let\P=\Pi     \let\Ps=\Psi   \let\Si=\Sigma
\let\Th=\Theta  \let\X=\Xi     \let\Y=\Upsilon

%

%

\def\cA{{\cal A}}                \def\cB{{\cal B}}
\def\cC{{\cal C}}                \def\cD{{\cal D}}
\def\cE{{\cal E}}                \def\cF{{\cal F}}
\def\cG{{\cal G}}                \def\cH{{\cal H}}
\def\cI{{\cal I}}                \def\cJ{{\cal J}}
\def\cK{{\cal K}}                \def\cL{{\cal L}}
\def\cM{{\cal M}}                \def\cN{{\cal N}}
\def\cO{{\cal O}}                \def\cP{{\cal P}}
\def\cQ{{\cal Q}}                \def\cR{{\cal R}}
\def\cS{{\cal S}}                \def\cT{{\cal T}}
\def\cU{{\cal U}}                \def\cV{{\cal V}}
\def\cW{{\cal W}}                \def\cX{{\cal X}}
\def\cY{{\cal Y}}                \def\cZ{{\cal Z}}
%

\newcommand{\Ns}{N\hspace{-4.7mm}\not\hspace{2.7mm}}
\newcommand{\qs}{q\hspace{-3.7mm}\not\hspace{3.4mm}}
\newcommand{\ps}{p\hspace{-3.3mm}\not\hspace{1.2mm}}
\newcommand{\ks}{k\hspace{-3.3mm}\not\hspace{1.2mm}}
\newcommand{\des}{\partial\hspace{-4.mm}\not\hspace{2.5mm}}
\newcommand{\desco}{D\hspace{-4mm}\not\hspace{2mm}}


%
\title{\boldmath
eV Seesaw with Four Generations }
\vfill
\author{Wei-Shu Hou$^{a}$}
\author{Andrea Soddu$^{b}$}
\affiliation{ $^a$Department of Physics, National Taiwan
 University, Taipei, Taiwan 10617, R.O.C. \\
$^b$Department of Particle Physics, Weizmann Institute
 of Science, Rehovot 76100, Israel
}
%

%
%
\vfill
\begin{abstract}
We extend the eV seesaw scenario to four lepton generations. The
LSND anomaly is taken as the right-handed seesaw scale, i.e. $m_R
\sim$ eV. The fourth generation then gives a heavy pseudo-Dirac
neutrino which largely decouples from other generations, and is
relatively stable. One effectively has a $3+3$ solution to the
LSND anomaly, where we illustrate with numerical solutions. Our
study seems to indicate that the third mixing angle
$\sin^2{\theta_{13}}$ may be less than 0.01.
\end{abstract}
%
\maketitle

\pagestyle{plain}

\section{Introduction}

Neutrino oscillation measurements have become more and more
accurate. The latest combined results \cite{Lisi} read $\D
m^2_{\odot} = (7.9 \pm 0.7) \times 10^{-5} \ {\rm eV}^2$and $\D
m^2_{\rm atm} = (2.4^{+0.5}_{-0.6})\times 10^{-3} \ {\rm eV}^2$,
for the mass squared differences of the large mixing angle
solution (LMA) to the solar neutrino problem \cite{solar1,solar2},
and for atmospheric neutrinos \cite{atm1,atm2,atm3}, respectively.
Both mass differences are sub-eV, but the neutrino mass scale is
not yet certain.

Neutrino data also hint at the possibility of more than three
massive, mostly active neutrinos. The Liquid Scintillator Neutrino
Detector (LSND) Collaboration \cite{LSND1} has reported evidence
for a $\bar{\nu}_e$ flux $30$ meters away from a source of
$\bar{\nu}_\mu$, produced in $\pi^+ \rightarrow \mu^+ \nu_\mu$
with subsequent $\mu^+ \rightarrow \bar{\nu}_\mu + e^+ + \nu_e$
decay. The unexpected flux can be explained if there is a small
probability $P(\bar{\nu}_\mu \rightarrow \bar{\nu}_e)=(0.26 \pm
0.08)\%$ for a neutrino produced as a $\bar{\nu}_\mu$ to be
detected as a $\bar{\nu}_e$ \cite{LSND2}.
This LSND anomaly has yet to be confirmed, and the MiniBooNE
experiment at Fermilab \cite{MiniBooNE} should very soon give a
definitive confirmation or refutation. Numerous attempts to solve
the LSND puzzle, however, have already been proposed
\cite{Strumia}. It has been shown that oscillations with extra
sterile neutrinos can fit the LSND anomaly \cite{sterile}. But it
has also been pointed out that extra sterile neutrinos could be in
conflict with Big Bang Nucleosynthesis (BBN) \cite{Cirelli}, as
well as SN1987A supernova neutrino events \cite{CPT}.

In a recent work \cite{Gouvea}, it has been argued that, if the
right-handed neutrino Majorana scale $m_R$ is of ${\cal O}({\rm
eV})$, adequate fits to the LSND data can be obtained. This ``eV
seesaw" scenario runs against theoretical arguments in favor of a
very large $m_R$. To name a few such arguments: the canonical
seesaw mechanism \cite{seesaw1,seesaw2,seesaw3} with $m_R \sim
10^{14}\ {\rm GeV}$ can elegantly explain why neutrino masses are
so small, even with lepton Yukawa couplings that are of order one;
thermal leptogenesis \cite{leptogenesis1} points to $m_R \gtrsim
10^{10}\ {\rm GeV}$ \cite{leptogenesis2}. However, as stressed in
Ref. \cite{Gouvea}, nothing {\it experimental} is really known
about the magnitude of $m_R$, except perhaps the LSND result,
which is at eV scale.

The purpose of this letter is to show that the eV seesaw proposed
in Ref.~\cite{Gouvea} can be straightforwardly incorporated in a
four generation scenario.

The Standard Model (SM) with a sequential fourth generation (SM4)
is not ruled out by electroweak precision measurements, if one
allows the extra active neutrino to have mass close to $50$ GeV
\cite{Maltoni,HPS}. To avoid bounds from direct search at LEP II
\cite{LEPII}, mixing of the fourth heavy neutrino with the three
light neutrinos should be small $(\lesssim 10^{-6})$. It is clear
that in standard seesaw with $m_R \sim 10^{15}$ GeV, an extra
generation is hard to accommodate (A different approach to predict
a light sterile neutrino in the presence of a fourth generation is
the so called ``flipped seesaw" \cite{GZ}). All four (mostly)
active neutrinos will then be light, contradicting the invisible
$Z$ width which measures only three light neutrinos. But taking
$m_R$ at scale ${\cal O}(\rm{eV})$, one can now have a
sufficiently heavy fourth neutrino.

In the following we will show that, by taking $m_R \sim {\cal
O}(\rm{eV})$, the fourth neutrino is pseudo-Dirac and heavy. It
will not affect the invisible $Z$ width, and largely decouples
from lower generations. Aside from three mostly active light
neutrinos, three sterile neutrinos with mass $\gtrsim {\rm eV}$ is
predicted. A numerical analysis gives results consistent with the
LSND as well as solar and atmospheric data. It seems that
$\sin^2\theta_{13}$ cannot be large.

\section{Pseudo-Dirac Fourth Neutrino}

Following Ref. \cite{Gouvea} but allowing for a possible 4th
generation, the $8 \times 8$ neutrino mass matrix $M$ is given by
\begin{equation}
M = M_D + \Delta M_R + \delta M_D,
 \label{massmatrix}
\end{equation}
in a form suggestive of mass hierarchies.
In the basis where the $4 \times 4$ Dirac mass matrix is diagonal,
the dominant Dirac mass for the 4th generation arises from
\begin{equation}
M_D = m_D\left(
\begin{array}{cc}
0 & I_4 \\
I_4 & 0
\end{array} \right),
\end{equation}
where $m_D \sim 50$ GeV, $0$ and $I_4$ are ${4\times4}$ matrices
with zero elements, except 1 in 44 element of $I_4$.
%
The right-handed Majorana mass matrix is given by
\begin{equation}
\Delta M_R=m_R \left(
\begin{array}{cc}
0 & 0 \\
0 & r  \\
\end{array} \right),
\label{MR}
\end{equation}
where $m_R \sim$ eV \cite{Gouvea}, and $r$ is a ${4\times4}$
symmetric matrix with elements $r_{ij} \sim 1$.
The third matrix is
\begin{equation}
\delta M_D = m_R \left(
\begin{array}{cc}
0 & \varepsilon \\
\varepsilon & 0
\end{array} \right),
\end{equation}
which is pinned more to the $m_R$ scale, with
\begin{equation}
{ \varepsilon =\left(
\begin{array}{cccc}
\ep_1 & 0 & 0 & 0  \\
0 & \ep_2 & 0 & 0  \\
0 & 0 & \ep_3 & 0  \\
0 & 0 &    0  & 0  \\
\end{array} \right)
},
\end{equation}
where $\ep_4$ has been absorbed into $m_D$. Clearly, with $m_R/m_D
\equiv x \sim 10^{-10}$ and $\ep_i$ considerably less than 1,
$\Delta M_R$ and $\delta M_D$ can be treated as perturbations to
$M_D$.

We note that the neutrino mass matrix of Eq.~(\ref{massmatrix})
could arise from very small deviations from a democratic structure
for the Dirac contribution \cite{Silva}, and lepton number is
assumed to be only slightly violated. The latter requires $m_R
\sim 0$ \cite{Gouvea}, i.e. symmetry is enhanced in the limit of
$\Delta M_R \rightarrow 0$. Having smaller elements in $\delta
M_D$ compared to $m_R$ is purely phenomenological \cite{Gouvea}.

The dominant $M_D$ has six null eigenvalues, plus two eigenvalues
$\pm m_D$. For the eigenvectors associated with zero eigenvalues,
one can choose $e_i^{(0)} = (0,\ldots,0,1,0\ldots,0)$ with $1$ in
the $i^{th}$ position for $i=1,2,3$ and $i=5,6,7$. For the
eigenvalues $-m_D$ and $m_D$, the corresponding eigenvectors are
$e_{4,8}^{(0)}=(0,0,0,\mp1/\sqrt{2},0,0,0,1/\sqrt{2})$.
%
At zeroth order in $x$, the states $e_4^{(0)}$ and $e_8^{(0)}$
combine into a pure Dirac state of mass $m_D$. When linear
corrections in $x$ are considered, the perturbed states $e_4$ and
$e_8$ with $e_i= e_i^{(0)} + x e_i^{(1)}$ have masses which differ
by ${\cal O}(x)$, and they now correspond to a pseudo-Dirac
neutrino with mass $\sim m_D$, which we denote as $N$ (the charged
partner is denoted $E$, with $m_E \gtrsim 100$ GeV
\cite{PDG2004}).

For the six null eigenvalues of $M_D$, one can apply perturbation
theory with degeneracies. A linear combination of the degenerate
unperturbed states $e_i^{(0)}$ diagonalizes the $\Delta M_R +
\delta M_D$ perturbation. That is,
by diagonalizing the $6\times6$ perturbation matrix
\begin{equation}
{\scriptsize M^{(3)}=m_R\left(
\begin{array}{cccccc}
0 & 0 & 0 & \ep_1 & 0 & 0 \\
0 & 0 & 0 & 0 & \ep_2 & 0  \\
0 & 0 & 0 & 0 & 0 & \ep_3 \\
\ep_1 & 0 & 0 & r_{11} & r_{12} & r_{13} \\
0 & \ep_2 & 0 & r_{21} & r_{22} & r_{23} \\
0 & 0 & \ep_3 & r_{31} & r_{32} & r_{33} \\
\end{array} \right)},
\label{M3x3}
\end{equation}
one obtains the corrections of ${\cal O}(x)$ to the mass
eigenvalues, and the correct eigenstates at order zero. For the
effect of the 4th generation neutrino through the right-handed
sector, $r_{ij}$, one has linear corrections in $x$ proportional
to $r_{44}$ to the eigenvalues $\pm m_D$ for $e_4$ and $e_8$ as
already stated, and ${\cal O}(x^2)$ corrections to all the other
eigenvalues. The big hierarchy between the matrix elements
$M_{48}$, $M_{84} \cong m_D$ and all the others allow the fourth
generation to largely decouple from the other three.
As stated in the Introduction, this is also required by direct
search limits that demand very small mixings between $N$ and the
light neutrino flavors.

Having reduced the problem to a $6\times 6$ case, the analysis
performed in \cite{Gouvea} suggests that one could find a solution
to the LSND puzzle. Our main goal here is to confirm the
possibility of an existing solution, and to gain some insight on
what could be a plausible scenario.

\

\section{3+3 Neutrino Model }

We set to zero all phases for simplicity, since there are already
too many parameters. We define $U^{\prime}$ to be the rotation
matrix which diagonalizes $M^{(3)}$. Having started in the basis
where the Dirac neutrino mass matrix $M_D + \delta M_D$ is
diagonal, one still has the freedom to perform a rotation
$U^{\prime \prime}$ in the left sector
\begin{equation}
{\scriptsize U^{\prime \prime}=\left(
\begin{array}{cccccc}
c_1 c_3 & s_1 c_3 & s_3 & 0 & 0 & 0 \\
-s_1 c_2 - c_1 s_2 s_3 & c_1 c_2 - s_1 s_2 s_3 & s_2 c_3 & 0 & 0 & 0  \\
s_1 s_2 - c_1 c_2 s_3 & -c_1 s_2 - s_1 c_2 s_3 & c_2 c_3 & 0 & 0 & 0 \\
0 & 0 & 0 & 1 & 0 & 0 \\
0 & 0 & 0 & 0 & 1 & 0 \\
0 & 0 & 0 & 0 & 0 & 1 \\
\end{array} \right)}.
\label{Uprpr}
\end{equation}
We assume no mixing between the fourth and first three generation
charged leptons, as already discussed.
For the right sector, a rotation will just change $r_{ij}$ to
$r_{ij}^{\prime}$, resulting in no change to our numerical
analysis.

The probability for a neutrino, produced with flavor $\alpha$ and
energy $E$, to be detected as a neutrino of flavor $\beta$ after
travelling a distance $L$ is \cite{PDG2004}
\begin{equation}
P(\nu_\a \rightarrow \nu_\b) = \delta_{\a \b}-4\sum_{j>i}^n U_{\a
j} U_{\b j}U_{\a i}U_{\b i} \, \sin^2{x_{ji}},
 \label{Palphabeta}
\end{equation}
where $\a=e, \mu, \tau, s_i$ with $s_i$ the sterile neutrino
flavors, $U=U^{\prime \prime} U^{\prime}$, and $x_{ji}=1.27\D
m_{ji}^2 \, L/E$ with $\D m_{ji}^2 \equiv m_j^2-m_i^2$.
Applying Eq.~(\ref{Palphabeta})
\cite{SCS} in the ``3 active plus 3 sterile neutrino"
 $(3+3)$ case, using the approximations $x_{21}=x_{31}=x_{32}=0$ and
$x_{i1}=x_{i2}=x_{i3}$ for $i=4,5,6$, one obtains
\begin{widetext}
\begin{eqnarray}
P(\nu_\a \rightarrow \nu_\b) &= & \d_{\a\b} +
 4[ U_{\a 4}^2(U_{\b 4}^2-\d_{\a\b})\sin^2{x_{41}} + U_{\a 5}^2(U_{\b 5}^2-\d_{\a\b}) \sin^2{x_{51}}
  + U_{\a 6}^2(U_{\b 6}^2 -\d_{\a\b}) \sin^2{x_{61}} \nonumber \\
&& + U_{\a 4}U_{\b 4}U_{\a 5}U_{\b
5}(\sin^2{x_{41}}+\sin^2{x_{51}} -\sin^2{x_{54}})
 + U_{\a 4}U_{\b 4}U_{\a 6}U_{\b 6}(\sin^2{x_{41}}+\sin^2{x_{61}}
 -\sin^2{x_{64}})\nonumber \\
 &&+ U_{\a 5}U_{\b 6}U_{\a 5}U_{\b 6}(\sin^2{x_{51}}+\sin^2{x_{61}} -\sin^2{x_{65}})]\, ,
 \label{Pmue}
\end{eqnarray}
\end{widetext}
where orthogonality of $U$ has been used.
Expressions for the mixing angles are given by \cite{Gouvea2}
\begin{eqnarray}
&&\tan^2{\theta_{12}}\equiv\frac{|U_{e2}|^2}{|U_{e1}|^2}\, , \,\,
\tan^2{\theta_{23}}\equiv\frac{|U_{\mu 3}|^2}{|U_{\tau 3}|^2} \, ,\nonumber \\
&& \ \sin^2{\theta_{13}}\equiv{|U_{e3}|^2}\, . \label{mixingangle}
\end{eqnarray}

\begin{table}
\caption{Best fit values, $2\s$ and  $3\s$  intervals for
three-flavor neutrino oscillation parameters from global data,
including solar, atmospheric, reactor (KamLAND and CHOOZ) and
accelerator (K2K) experiments, taken from Ref.~\cite{Maltoni2}.  }
\begin{center}
\begin{ruledtabular}
\begin{tabular}{cccc}
& BEST FIT & $2\s$ & $3\s$
\\ \hline \\
$\D m_{21}^2\,(10^{-5} \; {\rm eV}^2)$ & $8.1$  & $7.5-8.7$  & $7.2-9.1$    \\
$\D m_{31}^2\,(10^{-3} \; {\rm eV}^2)$ & $2.2$  & $1.7-2.9$  & $1.4-3.3$      \\
$\sin^2{\theta_{12}}$ & $0.30$  & $0.25-0.34$  & $0.23-0.38$    \\
$\sin^2{\theta_{23}}$ & $0.50$  & $0.38-0.64$  & $0.34-0.68$     \\
$\sin^2{\theta_{13}}$ & $0.0$  & $\leq 0.028$  & $\leq 0.047$      \\
\end{tabular}
\end{ruledtabular}
\end{center}
\end{table}

To perform our numerical analysis, we build the $\chi^2$ by using
the three-flavor neutrino oscillation parameters $\D m_{ji}^2$ and
$\sin^2{\theta_{ij}}$ taken from Ref.~\cite{Maltoni2}, which is
compiled from global data including solar, atmospheric, reactor
(KamLAND and CHOOZ) and accelerator (K2K) experiments. These are
given in Table 1. We also include the LSND result of
$P(\bar{\nu}_\mu \rightarrow \bar{\nu}_e)=(0.26\pm0.08)\%$, and
require $\D m^2_{41} \sim 1 \, {\rm eV}^2$, for a total of seven
inputs.
As one clearly has too many parameters to make a proper fit, we
just minimize the $\c^2$ built with these quantities by letting
the twelve parameters of the model vary.

As an illustration, we find $s_1=-0.57,\,s_2=0.98,\, s_3=0.80$,
which give
\begin{equation}
{\scriptsize
U^{\prime \prime}=\left(
\begin{array}{cccccc}
0.49 & -0.34 & 0.80 & 0 & 0 & 0 \\
-0.54 & 0.60 & 0.58 & 0 & 0 & 0  \\
-0.69 & -0.72 & 0.11 & 0 & 0 & 0 \\
0 & 0 & 0 & 1 & 0 & 0 \\
0 & 0 & 0 & 0 & 1 & 0 \\
0 & 0 & 0 & 0 & 0 & 1 \\
\end{array} \right) \, ,
} \label{Udoubleprime}
\end{equation}
and
%
\begin{equation}
{\scriptsize  M^{(3)}=5\,{\rm eV} \left(
\begin{array}{cccccc}
0 & 0 & 0 & 0.051 & 0 & 0 \\
0 & 0 & 0 & 0 & 0.004 & 0  \\
0 & 0 & 0 & 0 & 0 & 0.031 \\
0.051 & 0 & 0 & 1.195 & 0.861 & 1.038 \\
0 & 0.004 & 0 & 0.861 & 0.968 & 0.878 \\
0 & 0 & 0.031 & 1.038 & 0.878 & 1.264 \\
\end{array} \right),
} \label{M3x3num}
\end{equation}
%
from which, together with Eq.~(\ref{M3x3}), one can read the
values for the rest of the parameters coming from the minimization
process. The eigenvalues of $M^{(3)}$ are
\begin{eqnarray}
(m_1,\,m_2,\,m_3) &=& -(7 \times 10^{-5}, \, 9\times 10^{-3}, \,
0.048) \;{\rm eV},
\nonumber \\
(m_4, \, m_5, \, m_6) &=& (1, \, 1.2, \, 15) \; {\rm eV},
\label{masses}
\end{eqnarray}
and the associated rotation matrix is
\begin{equation}
{\scriptsize U^{\prime}=\left(
\begin{array}{cccccc}
-0.06 & -0.99 & -0.11 & 0 & 0 & 0 \\
-0.38 & 0.12 & -0.91 & 0.01 & -0.06 & 0.05  \\
0.90 & -0.02 & -0.37 & -0.17 & 0.05 & 0.12 \\
-0.19 & 0 & 0.09 & -0.75 & 0.16 & 0.60 \\
0.05 & -0.01 & 0.07 & 0.22 & -0.83 & 0.49 \\
0.01 & 0 & 0 & 0.60 & 0.52 & 0.61 \\
\end{array} \right).
} \label{Uprime}
\end{equation}

From Eq.~(\ref{masses}) one gets $\D m_{21}^2=8.1\times
10^{-5}\,{\rm eV}^2$ and $\D m_{31}^2=2.3\times 10^{-3}\,{\rm
eV}^2$, which of course is in good agreement with data. The
requirements for the neutrino mass splitting applied in
Ref.~\cite{SCS} for the $3+2$ case, $0.1\, {\rm eV}^2 \leq \D
m_{41}^2 \leq \D m_{51}^2\leq 100 \, {\rm eV}^2$, are also
satisfied. This guarantees that the approximations used to derive
Eq.~(\ref{Pmue}) are valid.

From Eqs.~(\ref{Udoubleprime}) and (\ref{Uprime}), we obtain the
full rotation matrix $U = U''U'$, i.e.
\begin{equation}
{\scriptsize
U=\left(
\begin{array}{cccccc}
0.82 & -0.54 & -0.04 & -0.14 & 0.06 & 0.08 \\
0.33 & 0.60 & -0.71 & -0.09 & -0.01 & 0.10  \\
0.42 & 0.59 & 0.69 & -0.03 & 0.05 & -0.03 \\
-0.19 & 0 & 0.09 & -0.75 & 0.16 & 0.60 \\
0.05 & -0.01 & 0.07 & 0.22 & -0.84 & 0.49 \\
0.01 & 0 & 0.01 & 0.60 & 0.52 & 0.61 \\
\end{array} \right).
} \label{U}
\end{equation}
Using $x_{ji}=1.27\D m_{ji}^2({\rm eV}^2) L({\rm m})/E({\rm MeV})$
with $L/E \sim 1$, together with Eq.~(\ref{Pmue}) for the $\mu
\rightarrow e$ case, Eqs.~(\ref{masses}) and~(\ref{U}), one
obtains $P(\bar{\nu}_\mu \rightarrow \bar{\nu}_e)=0.15\%$, which
is within 2$\sigma$ from the LSND central value.

$M^{(3)}$ in Eq.~(\ref{M3x3num}) can be viewed as deviating from
\begin{equation}
{\scriptsize M^{(3)} = m_R \left(
\begin{array}{cccccc}
0 & 0 & 0 & 0 & 0 & 0 \\
0 & 0 & 0 & 0 & 0 & 0  \\
0 & 0 & 0 & 0 & 0 & 0 \\
0 & 0 & 0 & 1 & 1 & 1 \\
0 & 0 & 0 & 1 & 1 & 1 \\
0 & 0 & 0 & 1 & 1 & 1 \\
\end{array} \right) },
\label{M3x3app}
\end{equation}
which has five zero eigenvalues and one nonzero eigenvalue equal
to $3m_R \sim 15$ eV, and diagonalized by
\begin{equation}
{\scriptsize U^{\prime }=\left(
\begin{array}{cccccc}
1 & 0 & 0 & 0 & 0 & 0 \\
0 & 1 & 0 & 0 & 0 & 0  \\
0 & 0 & 1 & 0 & 0 & 0 \\
0 & 0 & 0 & -\frac{1}{\sqrt{2}} & \frac{1}{\sqrt{2}} & 0 \\
0 & 0 & 0 & -\frac{1}{\sqrt{2}} & 0 & \frac{1}{\sqrt{2}} \\
0 & 0 & 0 & \frac{1}{\sqrt{3}} & \frac{1}{\sqrt{3}} & \frac{1}{\sqrt{3}} \\
\end{array} \right) }.
\end{equation}
Together with $U''$ of Eq.~(\ref{Uprpr}), one has
\begin{equation}
{\scriptsize U=\left(
\begin{array}{cccccc}
c_1 c_3 & s_1 c_3 & s_3 & 0 & 0 & 0 \\
-s_1 c_2 - c_1 s_2 s_3 & c_1 c_2 - s_1 s_2 s_3 & s_2 c_3 & 0 & 0 & 0  \\
s_1 s_2 - c_1 c_2 s_3 & -c_1 s_2 - s_1 c_2 s_3 & c_2 c_3 & 0 & 0 & 0 \\
0 & 0 & 0 & -\frac{1}{\sqrt{2}} & \frac{1}{\sqrt{2}} & 0 \\
0 & 0 & 0 & -\frac{1}{\sqrt{2}} & 0 & \frac{1}{\sqrt{2}} \\
0 & 0 & 0 & \frac{1}{\sqrt{3}} & \frac{1}{\sqrt{3}} & \frac{1}{\sqrt{3}} \\
\end{array} \right) }.
\label{Uapp}
\end{equation}

\begin{figure}[t]
  \centering
  \includegraphics[width=0.3\textwidth]{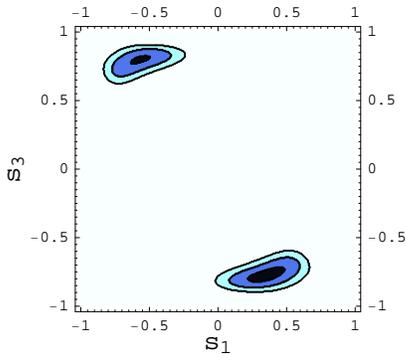}
  \vskip 0.5cm
  \caption{
  Contour-plot of $\chi^2$ vs the mixing angles $s_1$ and
  $s_3$, with $\ep_i$ and $r_{ij}$ as in
  Eq.~(\ref{M3x3num}), and $s_2 = 0.98$ held fixed.
  The regions in different shades are only indicative, and should
  not be interpreted as the $1\s$, $2\s$ and $3\s$ regions,
  as the rest of the parameters are fixed at the best fit values.
  }
  \label{fig:chiSQ23}
\end{figure}

\begin{figure}[b]
  \centering
  \includegraphics[width=0.45\textwidth,height=0.45\textwidth]{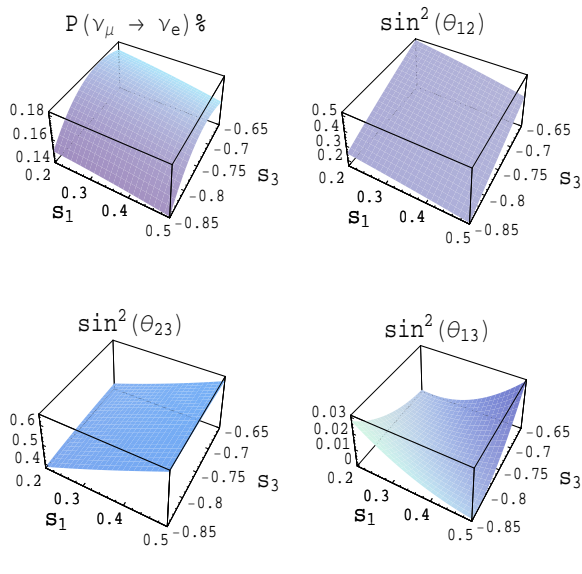}
  \caption{
  $P(\bar{\nu}_\mu \rightarrow \bar{\nu}_e)$, $\sin^2{\theta_{12}}$,
  $\sin^2{\theta_{23}}$ and $\sin^2{\theta_{13}}$ vs $s_1$ and $s_3$,
  corresponding to the lower right solution in Fig. 1, with
  $\ep_i$ and $r_{ij}$ fixed as in Eq.~(\ref{M3x3num}), and $s_2=0.98$.}
  \label{fig:allplot}
\end{figure}

\begin{figure}[b]
  \centering
  \includegraphics[width=0.45\textwidth,height=0.45\textwidth]{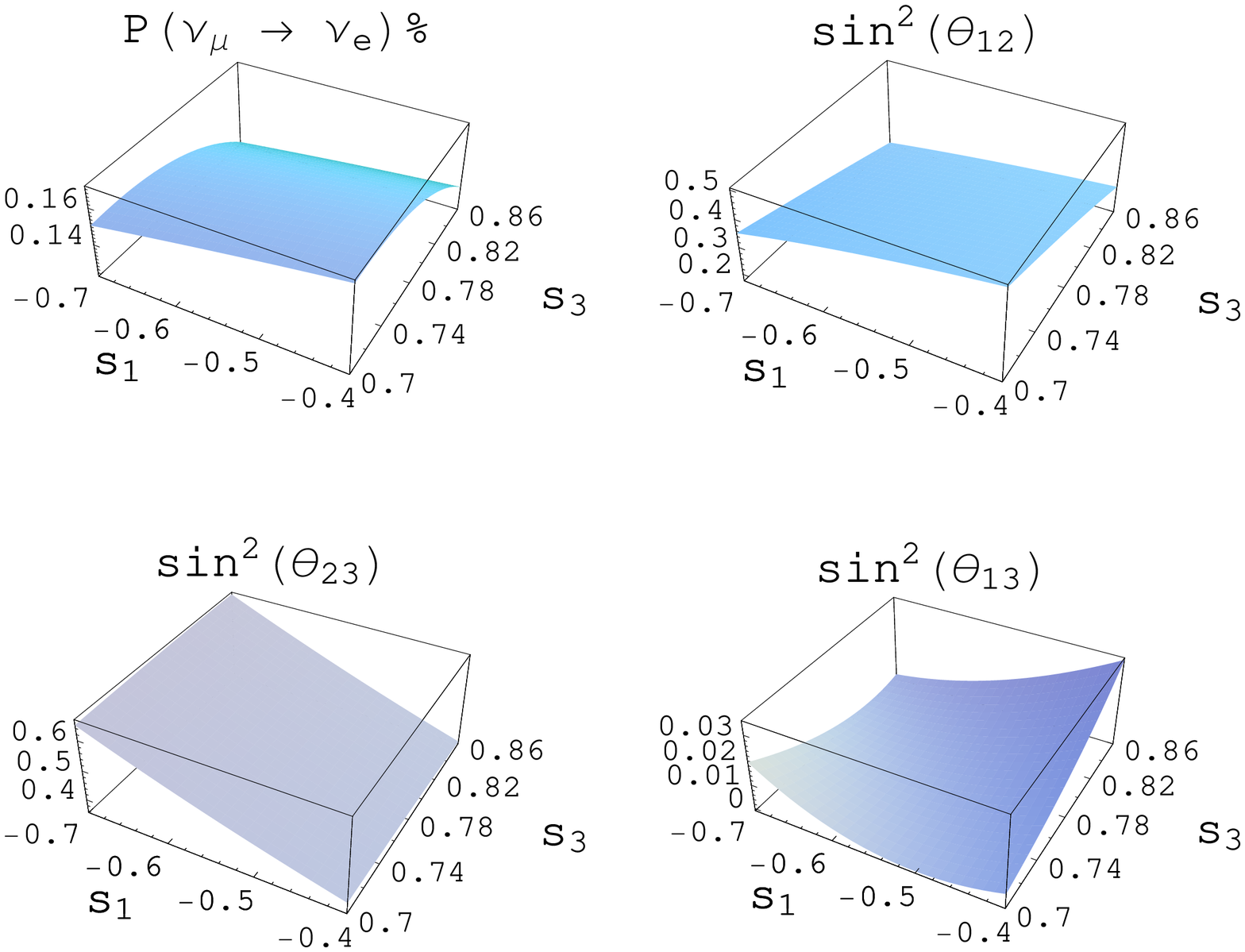}
  \caption{
  Same as Fig. 2, but corresponding to upper left solution in Fig. 1.}
  \label{fig:allplot}
\end{figure}

Applying this to Eqs.~(\ref{Pmue}) for the $\mu \rightarrow e$
case, one sees that the transition probability $P(\nu_\mu
\rightarrow \nu_e)$ would vanish. Deviations from
Eq.~(\ref{M3x3app}) as realized in Eq.~(\ref{M3x3num}) not only
produces the needed neutrino mass spectrum, finite transition
probability $P(\nu_\mu \rightarrow \nu_e)$ is also achieved.

In our analysis we did not take into account the null
short-baseline experiments (NSBL). Let us compare our results with
the ones obtained by doing a full analysis in Ref~\cite{SCS} for
the (3+2) case. Using the rotation matrix elements $U_{e4}=-0.14$,
$U_{\mu4}=-0.09$, $U_{e5}=0.06$, $U_{\mu5}=-0.01$, $U_{e6}=0.08$
and $U_{\mu6}=0.10$ as one can read from Eq.~(\ref{U}), together
with Eq.~(\ref{Pmue}) respectively for $\mu \rightarrow e$, $e
\rightarrow e$ and $\mu \rightarrow \mu$, we obtain $P(\nu_\mu
\rightarrow \nu_e)=0.0015$, $P(\nu_e \rightarrow \nu_e)=0.89$ and
$P(\nu_\mu \rightarrow \nu_\mu)=0.93$ in the $L/E \sim 1$
approximation. Our calculated values for oscillation appearance
and disappearance probabilities can now be compared with the ones
obtained by using the best fit values for the rotation matrix
elements in the (3+2) case as in Ref.~\cite{SCS}, $U_{e4}=0.121$,
$U_{\mu4}=0.204$, $U_{e5}=0.036$ and $U_{\mu5}=0.224$. In the $L/E
\sim 1$ approximation these give $P(\nu_\mu \rightarrow
\nu_e)=0.0021$, $P(\nu_e \rightarrow \nu_e)=0.95$ and $P(\nu_\mu
\rightarrow \nu_\mu)=0.84$. We remark that, although a full
analysis is needed to tell if our model is able to accommodate
both NSBL and LSND data, our predictions seem to be not too far
away from the results of Ref.~\cite{SCS}.


From Eqs.~(\ref{mixingangle}) and~(\ref{U}), we find
$\sin^2{\theta_{12}}=0.30$, $\sin^2{\theta_{23}}=0.52$, and
\begin{equation}
\sin^2{\theta_{13}}=0.0018. \label{theta13}
\end{equation}
As expected, the values for $\sin^2{\theta_{12}}$ and
$\sin^2{\theta_{23}}$ are in good agreement with data, but the as
yet unmeasured $\sin^2{\theta_{13}}$ turns out to be rather small.

\section{In Search of Sizable \boldmath $\sin^2{\theta_{13}}$}

To investigate the possibility for a bigger value of
$\sin^2{\theta_{13}}$, we restrict the $\chi^2$ to the four inputs
of $\sin^2{\theta_{ij}}$ and $P(\bar{\nu}_\mu \rightarrow
\bar{\nu}_e)$. The mass spectrum is not affected by the rotation
of Eq.~(\ref{Uprpr}). With $\ep_i$ and $r_{ij}$ given as in
Eq.~(\ref{M3x3num}), we first fix $s_1=-0.57$ and perform a
$\chi^2$ fit vs $s_2$ and $s_3$. We iterate with fixing $s_2 =
0.98$ ($s_3 = 0.8$) and minimize $\chi^2$ vs $s_1$ and $s_3$
($s_1$ and $s_2$). We find for both cases of fixing $s_1$ and
$s_3$ to the values found in previous section,
$\sin^2{\theta_{23}}$ is quite strongly dependent on $s_2$, and
the value around 0.98 is preferred. We thus illustrate with fixing
$s_2 = 0.98$.

In Fig. 1 we show the contour plot of $\chi^2$ vs $s_1, s_3$. The
three different shaded regions should not be interpreted as the
$1\s$, $2\s$ and $3\s$ regions, since we have fixed the rest of
the parameters to the best fit values. But they still give an
indication of variations around the best fit region under the
above assumptions.

In Fig. 2 we plot the four quantities $P(\bar{\nu}_\mu \rightarrow
\bar{\nu}_e)$, $\sin^2{\theta_{12}}$, $\sin^2{\theta_{23}}$ and
$\sin^2{\theta_{13}}$ vs $s_1$ and $s_3$, for the solution on the
lower right of Fig. 1. The same is plotted in Fig.~3 for the upper
left solution. Again, $\ep_i$ and $r_{ij}$ are fixed as in
Eq.~(\ref{M3x3num}), and $s_2$ is held fixed at 0.98. We see that
$P(\bar{\nu}_\mu \rightarrow \bar{\nu}_e)$ can reach the one
$\sigma$ region and $\sin^2{\theta_{12}}$ is well within range.
However, to push $\sin^2{\theta_{13}}$ beyond 0.01,
$\sin^2{\theta_{23}}$ seems to wander away from maximal mixing of
0.5, and values at $\sim 0.4$ or 0.6 has to be tolerated. We note
further that the sensitivity of $\sin^2{\theta_{23}}$ is to $s_1$,
rather than $s_3$.

We conclude that $\sin^2{\theta_{13}}$ greater than 0.01 is
possible, but seemingly not preferred. It is not clear whether
this is an artefact of not being able to do a real fit.
Note that we have not checked explicitly whether constraints from
short baseline disappearance experiments are fully satisfied.

\section{Discussion and Conclusion}

It is tempting to consider whether mixing between the fourth and
the first three light charged lepton generations could modify the
situation with $\sin^2{\theta_{13}}$. But as already mentioned in
the Introduction, one needs to satisfy both bounds from direct
search at LEP II and electroweak precision measurements. We have
pursued a numerical study, but find that, if we wish to keep the
mixings sufficiently small so that the fourth active heavy
neutrino will be semi-stable, no important change with respect to
the no-mixing case is observed. We note that the heavy neutrino
could be heavier than 50 GeV and still with suppressed mixing to
lower generations, but then one would have to face electroweak
precision constraints. We note in passing that semi-stable heavy
neutrinos are still of interest~\cite{Rybka} to dark matter search
experiments, as a fourth heavy lepton was once a leading dark
matter candidate.

In summary, we have extended the eV seesaw scenario to four lepton
generations. Taking the LSND scale as the right-handed seesaw
scale $m_R \sim$ eV, one has a heavy pseudo-Dirac neutrino with
mass $m_N \sim 50$ GeV, which largely decouples from other
generations, and is relatively stable. One effectively has a $3+3$
solution to the LSND anomaly, where we illustrate with numerical
solutions. As a possible outcome, our numerical study indicates
that the third mixing angle, $\sin^2{\theta_{13}}$, seems to be
less than 0.01.

\vskip 0.3cm \noindent{\bf Acknowledgement}.\ \ This work is
supported in part by NSC-94-2112-M-002-035 and HPRN-CT-2002-00292.
We thank G. Raz, T. Volansky and R. Volkas for useful discussions.

\end{document}